\documentclass[10pt,a4paper]{article}
\pdfoutput=1
\usepackage{jcappub}
\usepackage[english]{babel}
\usepackage{epsfig,graphicx}
\usepackage{amsfonts,amsmath,amssymb}
\usepackage{enumerate}
\usepackage{tabularx,varwidth,array,ragged2e,multirow}

\newcommand{\nc}{\newcommand}
\nc{\eq}[1]{Eq.~(\ref{#1})}
\nc{\eqs}[1]{Eqs.~(\ref{#1})}
\nc{\papref}[1]{[\ref{#1}]}

\nc{\pd}{\partial}
\nc{\bea}{\begin{eqnarray}}
\nc{\eea}{\end{eqnarray}}
\nc{\bal}{\begin{alignedat}}
\nc{\eal}{\end{alignedat}}
\nc{\beq}{\begin{equation}}
\nc{\eeq}{\end{equation}}
\nc{\bit}{\begin{itemize}}
\nc{\eit}{\end{itemize}}
\nc{\benu}{\begin{enumerate}}
\nc{\eenu}{\end{enumerate}}
\nc{\bdes}{\begin{description}}
\nc{\edes}{\end{description}}

\nc{\nn}{\nonumber}

\nc{\hc}{\text{h.c.}}
\nc{\cc}{\text{c.c.}}

\nc{\sub}[1]{_{\text{#1}}}
\nc{\ssub}[1]{_{_\text{#1}}}
\nc{\super}[1]{^{\text{#1}}}
\nc{\ssuper}[1]{^{^\text{#1}}}

\nc{\slashed}[1]{{#1}\hspace{-2mm}/}

\nc{\pare}[1]{\left( #1 \right)}
\nc{\sqpare}[1]{\left[ #1 \right]}
\nc{\ang}[1]{\left\langle #1 \right\rangle}
\nc{\abs}[1]{\left| #1 \right|}

\def\g5{\gamma_{5}}

\def\GeV{{\rm GeV}}
\def\TeV{{\rm TeV}}

\def \cm{{\rm cm}}
\def \km{{\rm km}}

\def \snd{{\rm s}}

\def\a{\alpha}

\def\g{\gamma}

\def\z{\zeta}

\def\th{\theta}

\def\m{\mu}
\def\n{\nu}

\def\p{\pi}
\def\r{\rho}
\def\s{\sigma}

\def\w{\omega}

\def\G{\Gamma}
\def\D{\Delta}

\def\W{\Omega}

\def\DM  {_{_{\rm DM}}}
\def\BSF {_{_{\rm BSF}}}

\def\mpl {M_{\rm Pl}}
\def\vrel {v_{\rm rel}}
\def\sann {\s_{\rm ann}}
\def\Sann {S_{\rm ann}}
\def\Sbarann {\bar{S}_{\rm ann}}
\def\sBSF {\s\BSF}
\def\SBSF {S\BSF}
\def\SbarBSF {\bar{S}\BSF}

\title{Bound-state formation  for thermal relic dark matter and unitarity}

\author[a]{Benedict von Harling}
\author[b]{and Kalliopi Petraki}

\affiliation[a]{SISSA and INFN, Via Bonomea 265, 34136 Trieste, Italy}
\affiliation[b]{Nikhef, Science Park 105, 1098 XG Amsterdam, The Netherlands}

\emailAdd{bharling@sissa.it}
\emailAdd{kpetraki@nikhef.nl}

\abstract{
We show that the relic abundance of thermal dark matter annihilating via a long-range interaction, is significantly affected by the formation and decay of dark matter bound states in the early universe, if the dark matter mass is above a few TeV.  We determine the coupling required to obtain the observed dark matter density, taking into account both the direct 2-to-2 annihilations and the formation of bound states, and provide an analytical fit. We argue that the unitarity limit on the inelastic cross-section is realized only if dark matter annihilates via a long-range interaction, and we determine the upper bound on the mass of thermal-relic dark matter to be about 197 (139)~TeV for (non)-self-conjugate dark matter.
}
\arxivnumber{1407.7874}

\begin{document}
\maketitle

\section{Introduction}

The annihilation cross-section of dark matter (DM) is largely what delimits the possibilities for its production and determines the expectations for its  phenomenology. In the standard paradigm, DM is assumed to have been in chemical equilibrium with the thermal bath in the early universe, due to rapid annihilation and pair-creation processes. As the universe expanded and cooled, these processes became inefficient, and the comoving DM density froze-out. In this scenario, the observed DM abundance determines the DM annihilation cross-section, provided that the DM particles are massive enough to have become non-relativistic at freeze-out. Weakly interacting massive particles naturally possess annihilation cross-sections in the vicinity of the required value~\cite{Chiu:1966kg,Lee:1977ua,Hut:1977zn,Wolfram:1978gp,Srednicki:1988ce,Kolb:1990vq,Gondolo:1990dk,Bertone:2004pz,Feng:2010gw}, and constitute the standard candidate for thermal-relic DM; however, thermal-relic DM may also reside in a hidden sector~\cite{Pospelov:2007mp}.

Here, we focus on heavy DM, with mass $m \gtrsim\,\TeV$. We discuss possibilities for the physics underlying the DM annihilation in this mass regime, and the couplings required to obtain the observed DM abundance from thermal freeze-out.  Heavy DM is of particular interest in view of the upcoming 14~TeV run of the LHC, as well as future high-energy experiments, such as a 100~TeV collider~\cite{Gershtein:2013iqa}. It is already being probed by direct and indirect detection experiments~\cite{Beltrame:2013bba,Akerib:2013tjd,Aartsen:2014gkd,FermiLAT:2011ab,Aguilar:2013qda}; in fact, DM with mass $m \gtrsim 500~\GeV$ has been invoked to explain the high-energy positron excess observed by PAMELA, Fermi and AMS~\cite{Adriani2008zr,FermiLAT:2011ab,Aguilar:2013qda,ArkaniHamed:2008qn,Pospelov:2008jd}. The precise knowledge of the DM couplings required for efficient annihilation in the early universe is essential in interpreting the experimental data.

The DM annihilation processes may be either due to short-range interactions mediated by heavy particles, as for example the weak interactions of the Standard Model (SM), or due to long-range interactions mediated by light species. The latter possibility becomes of interest typically when DM is hypothesized to reside in a hidden sector~\cite{Pospelov:2007mp}; however, even the weak interactions of the SM can manifest as long-range if DM is heavier than a few TeV~\cite{Hisano:2002fk,Hisano:2003ec}.  For $s$-wave annihilation via a short-range interaction,  the annihilation cross-section times relative velocity required to obtain the observed DM density from the freeze-out of thermal particles is  $\ang{\sann \vrel}_c \simeq 4.4 \times 10^{-26} \, \cm^3 \, \snd^{-1}$, assuming non-self-conjugate DM with mass above 10~GeV~\cite{Steigman:2012nb}.  If DM couples to a light force mediator, the long-range interaction between two incoming DM particles distorts their wavepackets; this is the well-known Sommerfeld effect~\cite{Sommerfeld:1931}. As a result, $\sann \vrel$ is enhanced at low velocities.  This Sommerfeld enhancement (SE) depends on the coupling strength of the DM to the light force mediator (c.f.~\eq{eq:S ann}).  The efficient annihilation of heavy DM in the early universe requires a large coupling, which renders SE significant during freeze-out. Indeed, the SE of the 2-to-2 annihilation processes affects the abundance of thermal-relic DM if the DM mass is $m \gtrsim 800~\GeV$~\cite{Hisano:2006nn,Feng:2009hw,Feng:2010zp,Hannestad:2010zt,Hryczuk:2010zi,Beneke:2014gja,Beneke:2014hja}; as a result, the coupling required to reproduce the observed DM density is lower than that estimated from $\ang{\sann \vrel}_c$ in the absence of SE.

Attractive long-range interactions imply also the existence of \emph{bound states}. Particle-antiparticle bound states, as well as bound states of self-conjugate identical particles, decay promptly into the force carrier particles; their formation thus contributes to the annihilation rate of their constituent species. In the early universe, the formation of DM bound states -- a process which is also enhanced at low velocities by the Sommerfeld effect -- can boost the DM annihilation, and affect the relation between the DM relic abundance and the DM couplings. 
In this paper, we investigate this effect. For concreteness, we consider fermionic DM coupled to a massless vector boson, a dark photon. We show that the formation of dark positronium-like states in the early universe and their subsequent decay into dark photons, affect the DM relic abundance for DM masses $m \gtrsim {\rm few}~\TeV$. We calculate the dark fine structure constant which yields the observed DM density, taking into account the formation of bound states, and their ionization and decay in the early universe.

The importance of considering  long-range interactions is underscored by unitarity.  It has long been shown that unitarity and the thermodynamics of the early universe set an upper bound on the mass of thermal-relic DM~\cite{Griest:1989wd}. Indeed, unitarity implies an upper limit on the inelastic DM self-interaction cross-section which decreases with increasing DM mass; for $s$-wave annihilation, in the non-relativistic regime, this is~\cite{Griest:1989wd}
\begin{equation}
( \s_{\rm inel} \vrel )_{\rm max} = \frac{4\pi }{ m^2 \vrel} \, ,
\label{eq:sigma unitarity}
\end{equation}
where $m$ is the DM mass and $\vrel$ is the relative velocity of the incoming DM particles.
The inelastic cross-section includes, of course, all the processes which may result in the annihilation of DM particles.   Notably, the dependence of $( \s_{\rm inel} \vrel )_{\rm max}$ on $\vrel$  implies that the maximum value of the inelastic cross-section can be realized if the DM particles interact via a light or massless force carrier, giving rise to the SE at low velocities exhibited by \eq{eq:sigma unitarity}.  In the following, we determine the unitarity bound on the mass of thermal DM, by employing Boltzmann equations to account properly for the late-time annihilations occurring as a result of the SE.  Close to the unitarity limit, bound-state formation (BSF) is the most efficient annihilation channel.

The effect of BSF on the DM relic abundance has been previously considered in Ref.~\cite{Feng:2009mn}, which found it to be negligible for DM masses $m \lesssim 10~\TeV$.  However, we find some discrepancies in the way the cross-section for radiative formation of positronium-like states was adapted in \cite{Feng:2009mn} from \cite{BetheSalpeter_QM}; this led to underestimating the effect.  Here we show that BSF affects the relic abundance of DM with mass above a few TeV. We determine the unitarity bound on the mass of thermal-relic DM, and discuss BSF in relation to unitarity.

\section{Cross-sections and rates}
\label{sec:rates}

We shall consider DM consisting of Dirac fermions $X$, coupled to a dark gauge force U(1)$_D$, via the Lagrangian
\beq
{\cal L} = \bar{X}(i\slashed{D}-m)X -\frac{1}{4} F_{\m\n} F^{\m\n} \, ,
\label{eq:Lagr}
\eeq
where $D^\m = \partial^\m + i g A^\m$, $F^{\m\n} = \partial^{\m}A^\n - \partial^{\n}A^\m$, with $A^\m$ being the dark photon field and  $\a \equiv g^2/(4\p)$ being the dark fine structure constant. 
As in QED, we will use $A^\m$ for the field in the Lagrangian, and $\g$ for the dark photon when discussing processes such as $X \bar{X} \to \g\g$. We omit the subscript $D$ for simplicity, as there is no risk of confusion with ordinary electromagnetism.

The direct annihilation of DM and the formation of DM bound states both contribute to the inelastic scattering of DM. Once they form, bound states may either decay into dark photons, or get ionized by the ambient radiation. The DM relic abundance depends on the balance of these processes. Below we list the pertinent cross-sections and rates.

\subsection{Annihilation, $X \bar{X} \to \g \g$}
In the non-relativistic regime, and to lowest order in $\a$, the 2-to-2 annihilation cross-section times relative velocity is $\s_0 \equiv (\s_{\rm ann} v_{\rm rel})_0 =  \pi \a^2 / m^2$~\cite{Dirac}.  Summing over the ladder diagrams involving photon exchange between $X$ and $\bar{X}$, yields the non-perturbative result
\beq 
\sann \vrel = \s_0 \: \Sann (\z) \, , \label{eq:sigma ann}
\eeq
where $\z \equiv \a / \vrel$, and $\Sann$ is the SE factor;
for $s$-wave annihilation (see e.g.~\cite{Cassel:2009wt,Iengo:2009ni})
\begin{equation}
\Sann (\z)  = \frac{2\p \z}{1- e^{-2\p \z}}  \,  .
\label{eq:S ann}
\end{equation}
At $\z \gtrsim 1$, $\Sann (\z) \simeq 2\p\z$. We show a plot of $\Sann (\z)$ in Fig.~\ref{fig:Sbar}.

\subsection{Bound-state formation, $X \bar{X} \to (X\bar{X})_{\rm bound} + \g$}
$X$ and $\bar{X}$ can bind into positronium-like states, the spin-singlet (para-) state, and the spin-triplet (ortho-) state, with masses $m_{\uparrow\downarrow}^{(n)} = m_{\uparrow\uparrow}^{(n)} = 2m -\D_n$, where  $\D_n = \m \a^2/ (2n^2)$ is the $n$-th level binding energy and $\m = m/2$ is the $X - \bar{X}$ reduced mass.

Bound states form via emission of a dark photon. The cross-section times relative velocity can be conveniently cast in the form 
\beq
\sBSF^{(n)} \vrel = \s_0 \: \SBSF^{(n)} (\z) \, ,
\label{eq:sBSF vrel}
\eeq
where~\cite{Lifshitz_RelativisticQM,BetheSalpeter_QM}\footnote{Reference~\cite{AkhiezerMerenkov_sigmaHydrogen} found $\sBSF$ to be larger by a factor of 2. This would enhance the effect of BSF on the DM freeze-out.}$^,$\footnote{We note that $\SBSF$ is not the BSF enhancement factor due to the Sommerfeld effect, as $\sBSF \vrel$ is not equal to $\s_0$ when the Sommerfeld effect is neglected.}
\begin{align}
\SBSF^{(1)} (\z)  
&= \frac{2^{10} \p}{3} \: \frac{\z^5}{(1+\z^2)^2}  \: \frac{e^{-4 \z {\rm arc}\cot \z}}{1-e^{-2\p \z}} \, ,
\label{eq:S BSF 1}
\\
\SBSF^{(n)} (\z)  &= (1/n) \, \SBSF^{(1)} (\z/n)  \, .
\label{eq:S BSF n}
\end{align}
At $\z \gg 1$, $\SBSF^{(1)} (\z)  \simeq 2^{10} \p\z/(3\exp 4)$. We show a plot of $\SBSF^{(1)} (\z)$ in Fig.~\ref{fig:Sbar}.f Clearly, at large $\z$, BSF becomes more efficient than DM annihilation into two photons, with $\SBSF^{(1)} (\z) / \Sann(\z) \simeq 3.1$. In fact, summing over $n$ yields 
\beq
\SBSF^{\rm tot} (\z)  
= \sum_{n=1}^\infty  \SBSF^{(n)} (\z)
\ \stackrel{\z/n \gg 1}{\longrightarrow}  \ 
\SBSF^{(1)} (\z) \sum_{n=1}^\infty  \frac{1}{n^2}  \,  ,
\label{eq:S BSF}
\eeq
which implies an overall enhancement factor for BSF from capture to excited states of at most $\sum_{n=1}^\infty  1/n^2 = \p^2/6 \simeq 1.6$.
Nevertheless, because the capture into the ground state dominates, and because the excited states are longer-lived, in the following we shall consider only the $n=1$ states.  Wherever $\sBSF, \SBSF$ and $\D$ appear without an index specifying the level, the $n=1$ state is implied.

Note that in the non-relativistic regime, neglecting the spin-orbit coupling, the BSF cross-section is independent of the  spin configuration of the incoming particles, which remains conserved in the process. $\sBSF$ is the cross-section for any such process. The spin-averaged cross-sections for the formation of para- and ortho-states are 
$\s_{_{\rm BSF, \uparrow\downarrow}} =  \sBSF/4$ and $\s_{_{\rm BSF, \uparrow\uparrow}} =3\sBSF /4$  respectively.

\subsection{Thermal average}
To estimate the effect of annihilations and BSF on the DM abundance, we need to average the respective rates over the momentum distribution of DM in the early universe. It will be convenient to define the time variables
\begin{equation}
x_{_{(X)}} \equiv \frac{m}{T_{_{(X)}} }
\quad \text{and} \quad
z_{_{(X)}} \equiv \frac{\D}{ T_{_{(X)}} } = \frac{\a^2 x_{_{(X)}}}{4}
\,  ,
\label{eq:x,z}
\end{equation}
where $T$ is the temperature of the dark plasma, and $T_{_X}$ is the temperature of the DM particles.  We discuss their relation Sec.~\ref{sec:timeline}.

\begin{figure}[t]
\centering
\includegraphics[width=0.45\textwidth]{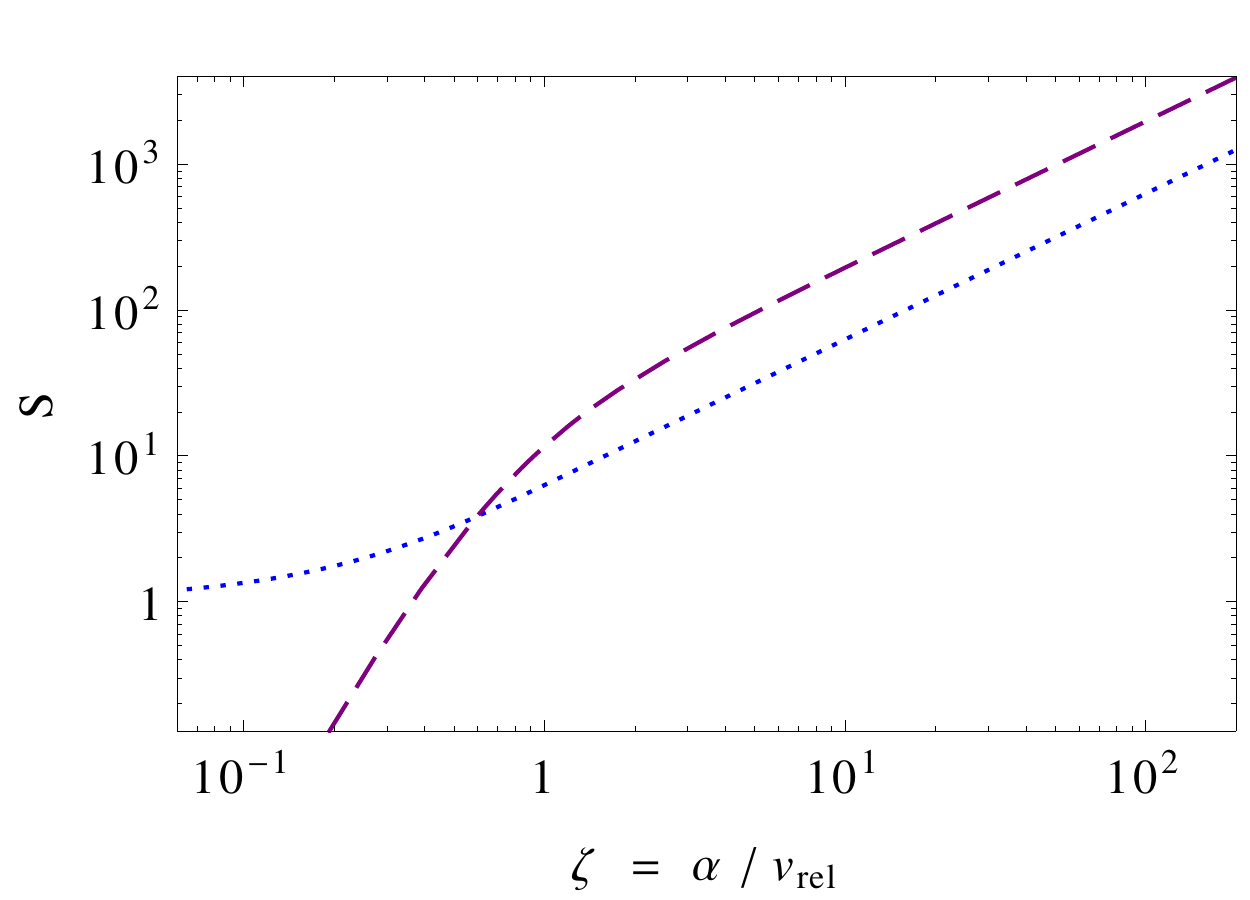}~~~
\includegraphics[width=0.45\textwidth]{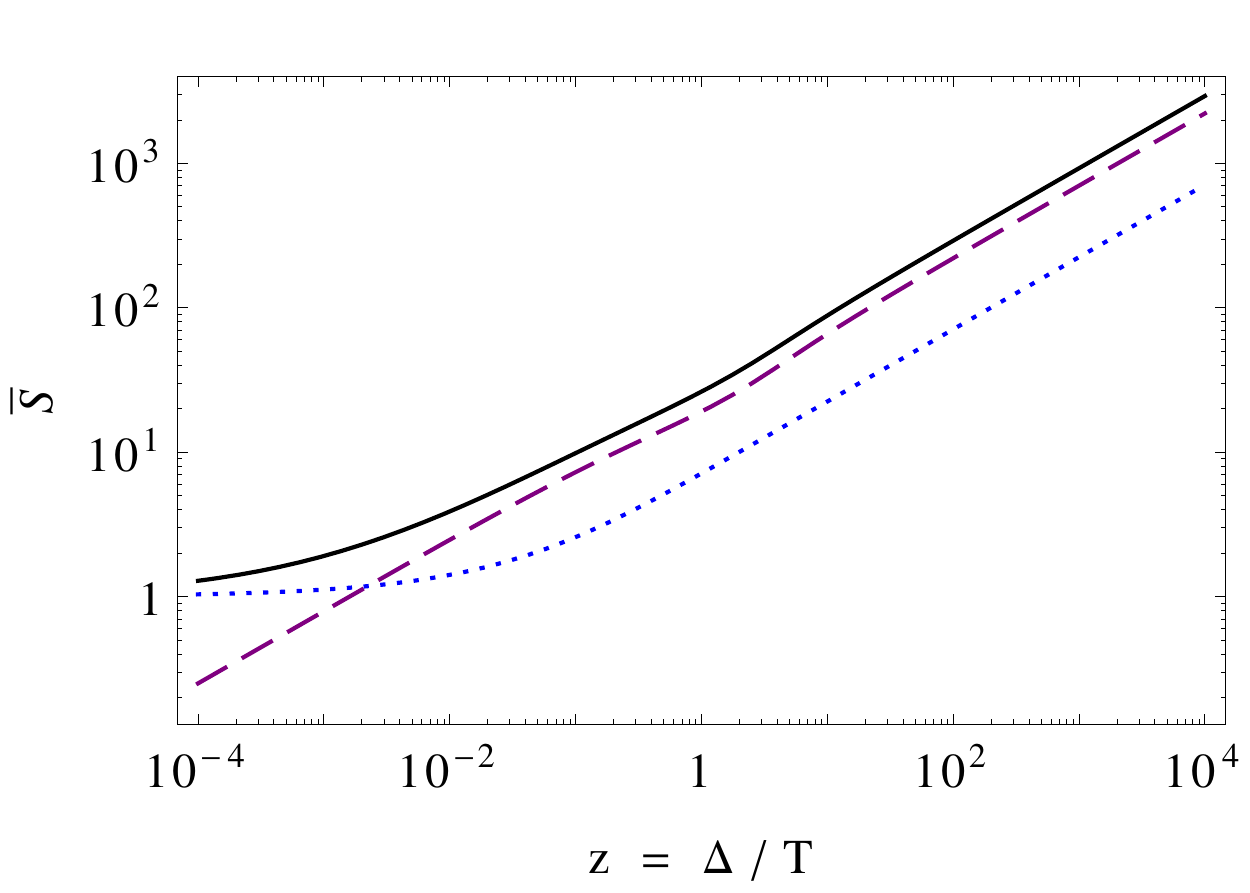}
\caption{
\emph{Left:} $S_{\rm ann}$ (dotted blue) and $S_{_{\rm BSF}}$ (dashed purple) vs. $\z = \alpha/\vrel$.
\emph{Right:} The thermally-averaged factors $\bar{S}_{\rm ann}$ (dotted blue) and $\bar{S}_{_{\rm BSF}}$ (dashed purple) vs.~$z=\Delta/T$, assuming $z=z_{_X}$. The solid black line is  $\bar{S}_{\rm tot} = \bar{S}_{\rm ann} + \bar{S}_{_{\rm BSF}}$.}
\label{fig:Sbar}
\end{figure}

Assuming a Maxwellian velocity distribution for the DM particles, the thermally averaged annihilation cross-section times relative velocity is 
\beq 
\ang{ \sann \vrel} = \s_0 \Sbarann(z_{_X}) \, , 
\eeq 
where~\cite{Feng:2010zp}
\bea
\Sbarann(z_{_X}) 
&=& \frac{x_{_X}^{3/2}}{2 \sqrt{\p}}  \int_0^\infty d\vrel \: \Sann (\a/\vrel) \: \vrel^2 \: e^{- x_{_X} \vrel^2/4} \nn \\
&=& \frac{2}{\sqrt{\p}} \int_0^\infty du \, \Sann \pare{\sqrt{z_{_X}/u}} \sqrt{u}   \, e^{- u}
\,  .
\label{eq:Sann bar}
\eea
For BSF, we include, for completeness, the Bose enhancement due to the final-state dark photon, which is emitted with energy $\w \simeq \D + \m \vrel^2/2 $.   The Bose enhancement remains important for $T\gtrsim \w$, i.e. typically until after freeze-out, though during this time the ionization of bound states is still rapid and impedes efficient DM annihilation via BSF (see below).  The BSF rate is proportional to  
\beq 
\ang{ \sBSF \vrel [1+f_\g(\w)] } = \s_0 \SbarBSF (z_{_X},z) \, , \label{eq:sigma BSF bar}
\eeq 
where $f_\g(\w) = 1/[\exp(\w/T)-1]$ and
\begin{equation}
\SbarBSF (z_{_X}, z) = 
\frac{2}{\sqrt{\p}} \int_0^\infty \!\! du \, \SBSF \! \pare{\! \sqrt{z_{_X}/u} }    
\, \frac{\sqrt{u} \, e^{z}}{ e^{z+u} - 1}  \,  .
\label{eq:Sbar BSF}
\end{equation}
At $z, z_{_X}\gg1$, $\SbarBSF \simeq 2^{11}\sqrt{\p z_{_X}} / (3 \exp 4)$ and $\Sbarann\simeq 4\sqrt{\p z_{_X}}$.  $\Sbarann$ and $\SbarBSF$ are shown in Fig.~\ref{fig:Sbar}.

\subsection{Decay of bound states}
The dark positronium-like states are unstable; the spin-singlet and spin-triplet states decay into two and three dark photons respectively. The corresponding decay rates are
\beq 
\G_{\uparrow \downarrow} = \a^5 \m 
\quad {\rm and} \quad 
\G_{\uparrow \uparrow} =  c_\a \: \a^5 \m \, , 
\label{eq:decay rates} 
\eeq
where $c_\a \equiv  4(\p^2 - 9) \a /(9\p)  \simeq 0.12 \a$~\cite{Stroscio1975215}.

\subsection{Ionization, $(X\bar{X})_{\rm bound} + \g \to X \bar{X}$}
The ionization cross-section of the bound states, $\s_{\rm ion}$, is related to $\sBSF$ by the Milne relation~\cite{Lifshitz_RelativisticQM}  
\beq 
\frac{\s_{\rm ion} }{\sBSF}  = \frac{\m^2 \vrel^2}{2\w^2} \, , 
\label{eq:Milne} 
\eeq 
where the factor of 2 counts the photon polarizations and $\w \simeq \D + \m\vrel^2/2$ is the photon energy.  The thermally averaged ionization rate is
\begin{equation}
\G_{\rm ion} (z) 
= \frac{2 \cdot 4\p}{(2\p)^3} \int_\Delta^\infty  \frac{d\w \, \w^2}{e^{\w/T} - 1} \, \s_{\rm ion} 
= \a^5 \m \: f_{\rm ion} (z) \,  ,
\end{equation}
where
\begin{equation}
f_{\rm ion} (z) \equiv
\frac{2^7}{3} \int_0^\infty \frac{d\eta \: \eta}{(1+\eta^2)^2} 
\frac{e^{-4\eta \: {\rm arc}\cot \eta}}{1-e^{-2\p\eta}}
\frac{1}{e^{z(1+1/\eta^2)}-1} \, . 
\end{equation}

\section{Timeline}
\label{sec:timeline}

\begin{table}
\centering
\def\arraystretch{1.3}
\begin{tabular}{ |l l l| }
\hline
$ z = z_f = x_f (\a^2/4)$ 			&\hspace{3mm} & freeze-out \\
$ z  \gtrsim 1.6 \times 10^{-3}$   	&& $\SbarBSF \geqslant 1$   \\
$ z  \gtrsim 2.2 \times 10^{-3}$		&& $\SbarBSF \geqslant \Sbarann $  \\
$ z  \gtrsim z_{\uparrow\downarrow} \simeq 0.28$   	   			&& $\G_{\uparrow\downarrow} \geqslant \G_{\rm ion} $   \\
$ c_\a \geqslant f_{\rm ion}(z_{\uparrow\uparrow})$  	&& $\G_{\uparrow\uparrow} \geqslant \G_{\rm ion} $  \\
$ z \gtrsim z_{\rm kd}$ 				&& kinetic decoupling 
\\ \hline
\end{tabular}
\caption{Timeline.}
\label{tab:timeline}
\end{table}

DM remains in chemical equilibrium with the dark photons due to annihilation, BSF and the inverse processes. As usual, we define the freeze-out of these processes as the time when the DM abundance differs from the equilibrium value by a factor of order 1. We estimate $x=x_f$ at freeze-out by adapting the standard result~\cite{Kolb:1990vq} to incorporate the SE of $\sann$ and $\sBSF$,
\beq
x_f + \frac{1}{2} \ln x_f - \ln \sqpare{\Sbarann (z_f) + \SbarBSF(z_f,z_f) }  
\approx  \ln \sqpare{ 0.038 \, (g_{_X}/\sqrt{g_*}) \, m \, \s_0 \mpl  } \, ,
\label{eq:xf}
\eeq
where $z_f = \a^2 x_f/4$, $g_{_X} = 2$ are the degrees of freedom of the DM species (the antiparticles are counted separately), and  $g_*$ is the number of relativistic degrees of freedom. Typically, $x_f \sim 25$.  Because of the SE, DM annihilations (direct and via BSF) continue to be significant after freeze-out.  To account for this, we integrate the Boltzmann equations for $z> z_f$, as described below.

After chemical decoupling, the DM particles remain in kinetic equilibrium with the dark photons via Thomson scattering, typically until rather late. The kinetic decoupling of the DM from the dark radiation occurs at (see e.g. Ref.~\cite{Feng:2010zp}) 
\beq 
z = z_{\rm kd} \sim 10^2 \pare{\frac{\a}{0.02}}^3 \pare{\frac{\TeV}{m}}^{1/2} . 
\label{eq:zkd} 
\eeq  
At $z \lesssim z_{\rm kd}$, $z_{_X}=z$.  After kinetic decoupling, provided that the effect of any residual interactions is negligible, 
$z_{_X} = z^2/z_{\rm kd}$~\cite{Kolb:1990vq}.

Here, we assume that the dark photons are at the same temperature as the plasma of SM particles.  This is a viable possibility during the freeze-out of DM with mass $m \gtrsim 100~\GeV$. Indeed, DM then freezes-out at a temperature $T_f = m/x_f \sim m/25 \gtrsim 4~\GeV$, before the QCD phase transition. Provided that the dark radiation has decoupled from the SM at that time, the subsequent decoupling of the QCD degrees of freedom reheats the SM plasma, and leads to the dark radiation having a lower temperature than ordinary photons; this ensures that the BBN and CMB constraints on the total relativistic energy of the universe are satisfied.  It is, of course, straightforward to generalize our calculation to the case that the dark radiation bath and the SM plasma are at different temperatures during DM freeze-out.

\begin{figure}[t]
\centering
\includegraphics[width=9cm]{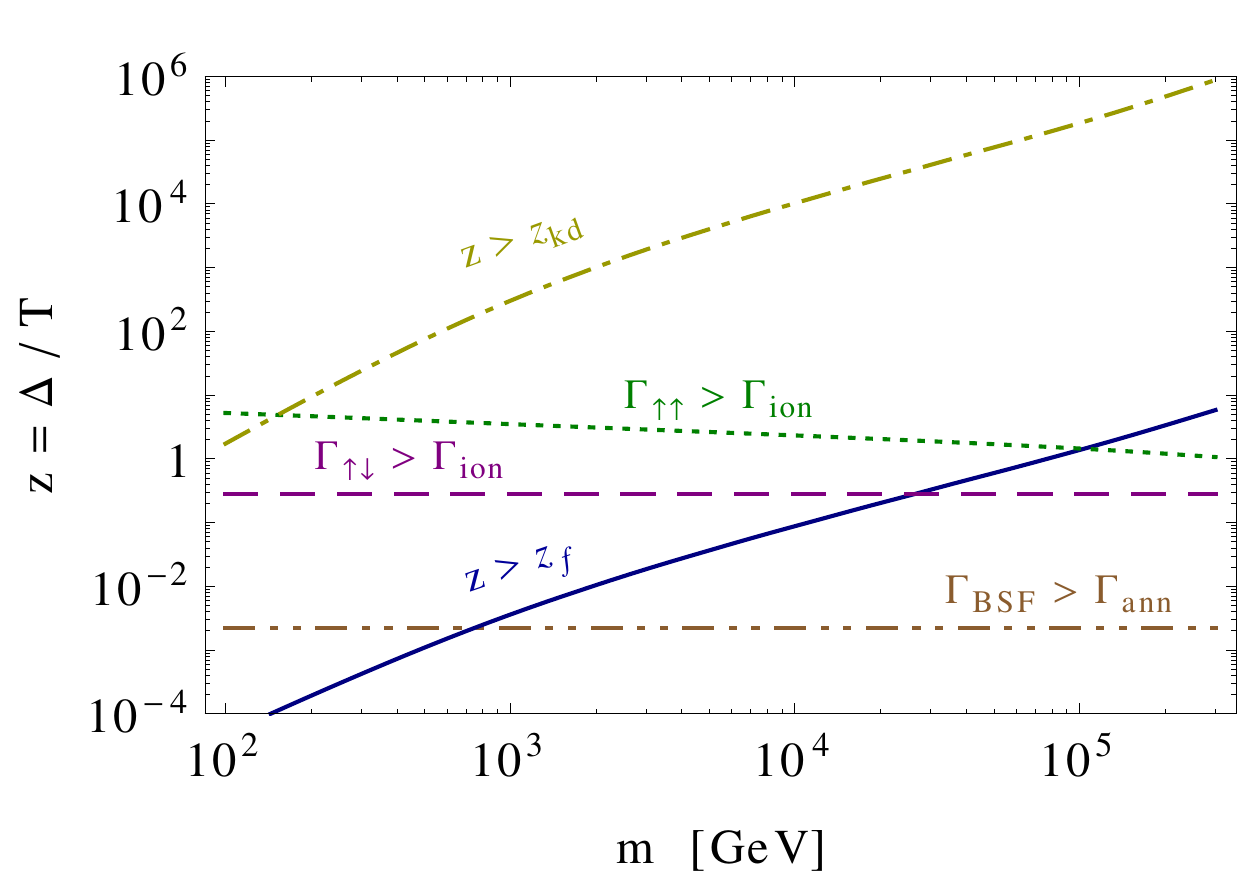}
\caption{$z = \Delta/T$ vs. the DM mass $m$: At the freeze-out time, $z=z_f$ (blue solid); the time when the rate of BSF exceeds the annihilation rate (brown dot-dot-dashed); the time when the decay of the dark para-positronium becomes faster than ionization, $z = z_{\uparrow\downarrow}$ (purple dashed); the time when the decay of the dark ortho-positronium becomes faster than ionization, $z = z_{\uparrow\uparrow}$ (green dotted); the time of kinetic decoupling, $z = z_{\rm kd}$ (yellow dot-dashed). We have used the values of $\alpha = \alpha(m)$ which reproduce the observed DM density. 
We may observe that at the time of freeze-out, BSF is faster than 2-to-2 annihilation if $m \gtrsim 753~{\rm GeV}$, dark para-positronium decay is faster than ionisation if $m \gtrsim 27~{\rm TeV}$, and dark ortho-positronium decay is faster than ionisation if $m\gtrsim 104~{\rm TeV}$.}
\label{fig:zVm}
\end{figure}

Based on the cross-sections and rates given above, in table~\ref{tab:timeline} we list (in approximate chronological order) the various mileposts which affect the DM relic abundance. In Fig.~\ref{fig:zVm}, we sketch these important times as functions of the DM mass, 
using the values of $\a$ which reproduce the observed DM abundance, as calculated in the next section (c.f.~Fig.~\ref{fig:alphaVm}).  Importantly, the efficiency of BSF in annihilating the thermal population of DM depends not only on $\sBSF$, but also on the balance between the decay and the ionization of the bound states which are formed.  
For $m \gtrsim 25.5~\TeV$, $z_{\uparrow\downarrow} \lesssim z_f$, and for $m\gtrsim 101.6~\TeV$, $z_{\uparrow\uparrow} \lesssim z_f$. This means that in the corresponding mass ranges, the DM annihilation via formation and decay of dark para- and ortho-positronium respectively, is significant already before freeze-out. 

\section{Relic abundance}

\subsection{Boltzmann equations}
\begin{figure}[t!]
\centering
\includegraphics[width=9cm]{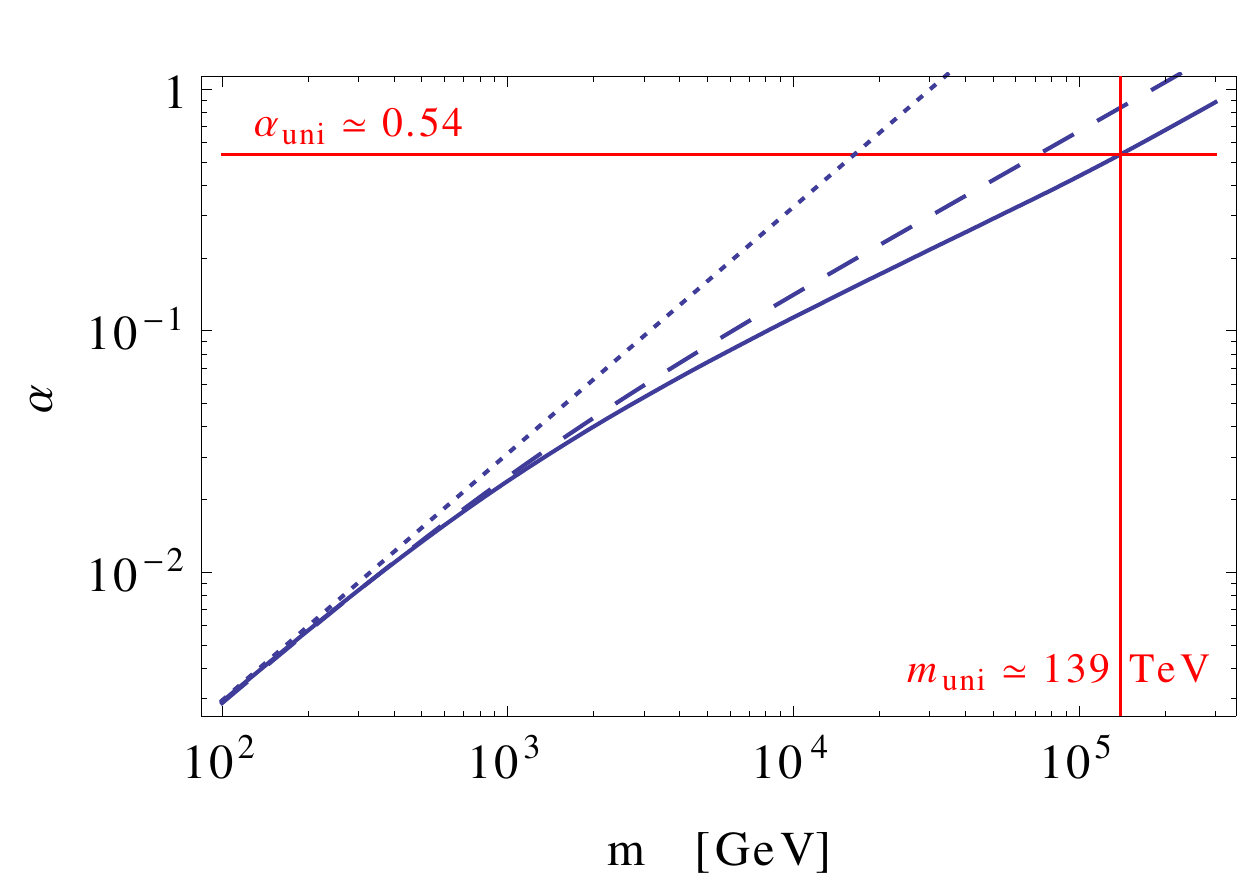}
\caption{
The dark fine-structure constant which reproduces the observed DM abundance from the thermal freeze-out of Dirac fermions vs. the DM mass.  
The solid blue line incorporates the effect of BSF and the SE of the direct DM annihilation into two dark photons. The dashed blue line neglects BSF, while the dotted blue line neglects both BSF and the SE of the 2-to-2 annihilation. The vertical and the horizontal solid red lines mark respectively the unitarity bound on the mass of thermal-relic DM and the value of $\alpha$ at which this is reached (evaluated assuming BSF into the ground state only). 
}
\label{fig:alphaVm}
\end{figure}
Let $Y_{_X} \equiv n_{_X}/s$, $Y_{\uparrow\downarrow} \equiv n_{\uparrow\downarrow}/s$ and $Y_{\uparrow\uparrow} \equiv n_{\uparrow\uparrow}/s$, where $n_{_X}, n_{\uparrow\downarrow}$ and $n_{\uparrow\uparrow}$ are the number densities of the unbound $X$ particles, the para- and the ortho-bound states respectively. $s = (2\p^2/45)g_{*S} T^3$ is the entropy density of the universe, with $g_{*S}$ being the entropic relativistic degrees of freedom. The abundances of the unbound and bound DM particles are governed by the coupled Boltzmann equations
\begin{align}
\frac{d Y_{_X}}{dz} = \:
& - \frac{c_1 \Sbarann(z_{_X})}{z^2} \: (Y_{_X}^2 - Y_{\rm eq}^2)   
  - \frac{c_1 \SbarBSF (z_{_X},z) }{z^2} \: Y_{_X}^2   
  + c_2  \, z \, f_{\rm ion}(z) \, ( Y_{\uparrow\downarrow} + Y_{\uparrow\uparrow}) \, ,
\label{eq:Boltz X} 
\\
\frac{d Y_{\uparrow\downarrow}}{dz} = \:
& \frac{c_1 \SbarBSF (z_{_X},z)}{4z^2}  \: Y_{_X}^2
-  c_2  z \sqpare{ 1+f_{\rm ion}(z) } \, Y_{\uparrow\downarrow} \, ,
\label{eq:Boltz para}
\\
\frac{d Y_{\uparrow\uparrow}}{dz} = \:
& \frac{3c_1 \SbarBSF (z_{_X},z)}{4z^2} \: Y_{_X}^2  
-  c_2 z \sqpare{ c_\a + f_{\rm ion}(z) } \, Y_{\uparrow\uparrow} \, ,
\label{eq:Boltz ortho}
\end{align}
where 

\begin{align}
c_1 &\equiv \sqrt{\frac{\p}{45}} \mpl \, \D \, \s_0 \, \pare{ \frac{g_{*S}}{\sqrt{g_*}} }  \, ,
\label{eq:c1}
\\
c_2 &\equiv \sqrt{\frac{45}{4\p^3 {g_*}}} \frac{\mpl}{\D^2} (\a^5 \m) \, ,
\label{eq:c2}
\end{align}
and 
\beq 
Y_{\rm eq} (x)  \equiv \frac{n_{_X}^{\rm eq} }{s} 
=  \frac{90}{(2\p)^{7/2}} \frac{g_{_X}}{g_{*S}} \, x^{3/2} \, e^{-x}
\label{eq:Yeq}
\eeq
is the equilibrium number density of the $X$ particles normalized to $s$. We take $g_* = g_{*S} = 108.75$ to account for the SM plus the two dark-photon degrees of freedom, and assume that $g_*$ and $g_{*S}$ remain constant.

We numerically integrate \eqs{eq:Boltz X} -- \eqref{eq:Boltz ortho}, starting from $z = z_i =x_i (\a^2/4)$ with $x_i=5$, until $z = z_s =100 \, z_{\rm kd}$, using the initial conditions
\begin{align}
Y_{_X} (z_i)  &\: =\:  Y_{\rm eq} (x_i) \, , \\
Y_{\uparrow\downarrow} (z_i) &\: = \:  (1/3) Y_{\uparrow\uparrow} (z_i)
= (1/g_{_X}) \,Y_{\rm eq} (2x_i-z_i) \, . 
\label{eq:initial cond}
\end{align}
Increasing $z_i$ up to $z_f$ and varying $z_s$ within reasonable limits have a negligible effect. 
The thermal equilibrium values for $Y_{\uparrow\downarrow}(z_i)$ and $Y_{\uparrow\uparrow}(z_i)$ are warranted because BSF gets into equilibrium before freeze-out for the couplings of interest; 
nevertheless, choosing instead vanishing initial values does not appreciably change the result. We have checked that the BSF and ionization rates appearing in Eqs.~\eqref{eq:Boltz X} -- \eqref{eq:Boltz ortho} cancel each other when $Y_{_X}$, $Y_{\uparrow\downarrow}$ and $Y_{\uparrow\uparrow}$ are equal to their equilibrium values. The fractional DM relic density is 
\beq 
\W_{_X} = 2 m Y_{_X} (z_s) s_0 / \r_c \, , 
\label{eq:Omega} 
\eeq  
where the factor of 2 accounts for the sum of $X$ and $\bar{X}$. $\r_c \simeq 4.9 \times 10^{-6}~\GeV~\cm^{-3}$ and $s_0 \simeq 2795~\cm^{-3}$~\cite{Ade:2013zuv} are the critical energy density and the entropy density of the universe today.

We evaluate $Y_{_X} (z_s)$,  and determine $\a$, such that the observed DM density, $\W_{_X} = \W\DM \simeq 0.26$~\cite{Ade:2013zuv}, is reproduced.  We present $\a$ vs. $m$ in Fig.~\ref{fig:alphaVm},  where we compare it with the values of $\a$ obtained by neglecting BSF, and  those obtained by neglecting both BSF and the SE of the 2-to-2 annihilations.  As can be seen, the effect of BSF is significant for $m \gtrsim$~few~TeV.  We find that for $m \gtrsim 100~\GeV$ and up the unitarity bound (see below), the fit 
\beq 
\a = \a_0 \, \frac{m}{m_0} \, \sqpare{\frac{2}{1+(m/m_0)^r}}^s , 
\label{eq:fit} 
\eeq 
with $\alpha_0 = 0.0247$, $m_0 = 1.04$~TeV, $r = 1.28$, $s = 0.328$, reproduces the values of $\alpha$ found numerically to better than 1\% accuracy.

\begin{figure}[t!]
\centering
\includegraphics[width=9cm]{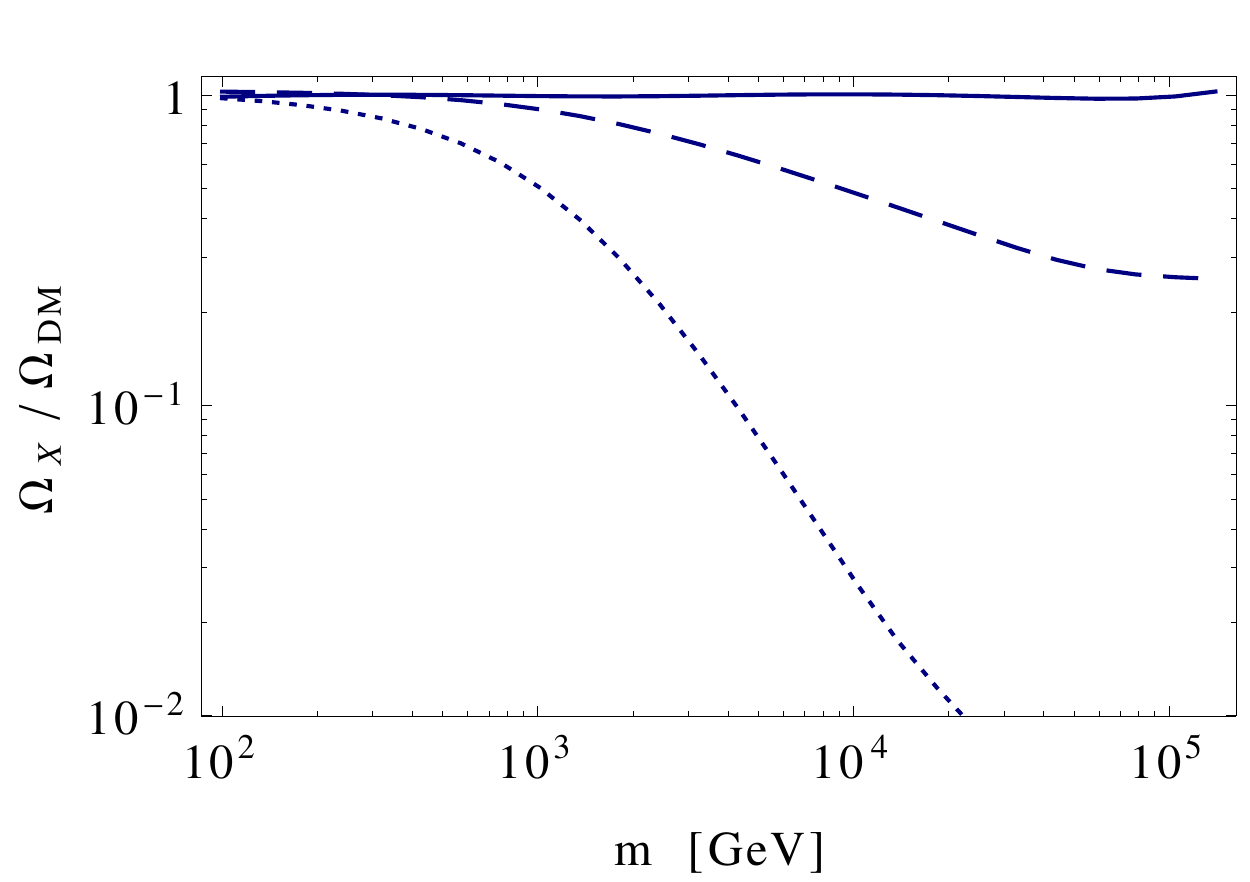}
\caption{The ratio of the relic density to the observed DM density $\Omega_{_X}/\Omega_{_{\rm DM}}$ vs. the DM mass $m$. The relic density $\Omega_{_X}$ is calculated by taking into account the Sommerfeld-enhanced annihilation and BSF. For the dark fine-structure constant $\alpha = \alpha(m)$, the values shown in Fig.~\ref{fig:alphaVm} are used, which assume that the following processes take place during the DM freeze-out: (i) annihilation without any Sommerfeld enhancement (dotted), (ii) Sommerfeld-enhanced annihilation only (dashed), (iii) Sommerfeld-enhanced  annihilation and BSF (solid). For the latter case, we have used the numerical fit of Eq.~\eqref{eq:fit}. From the dashed curve, we see that for $m \gtrsim$~few~TeV, the effect of BSF on the relic abundance exceeds the experimental uncertainty of about 1\% on the DM density. Similarly, for $m \gtrsim 10$~TeV, taking into account BSF yields a relic abundance which is less than half of what would have been calculated ignoring BSF.
}
\label{fig:Omegas}
\end{figure}

In Fig.~\ref{fig:Omegas}, we show the effect of BSF on the relic density of our DM candidate. Already for $m \gtrsim$~few~TeV, the effect of BSF on the relic abundance is larger than the uncertainty of about 1\% in the measurement of the DM density.
For $m \gtrsim 10~\TeV$, BSF diminishes the relic abundance to less than half of the value estimated when neglecting BSF.

\subsection{Effective Sommerfeld enhancement}

\begin{figure}[t!]
\centering
\includegraphics[width=9cm]{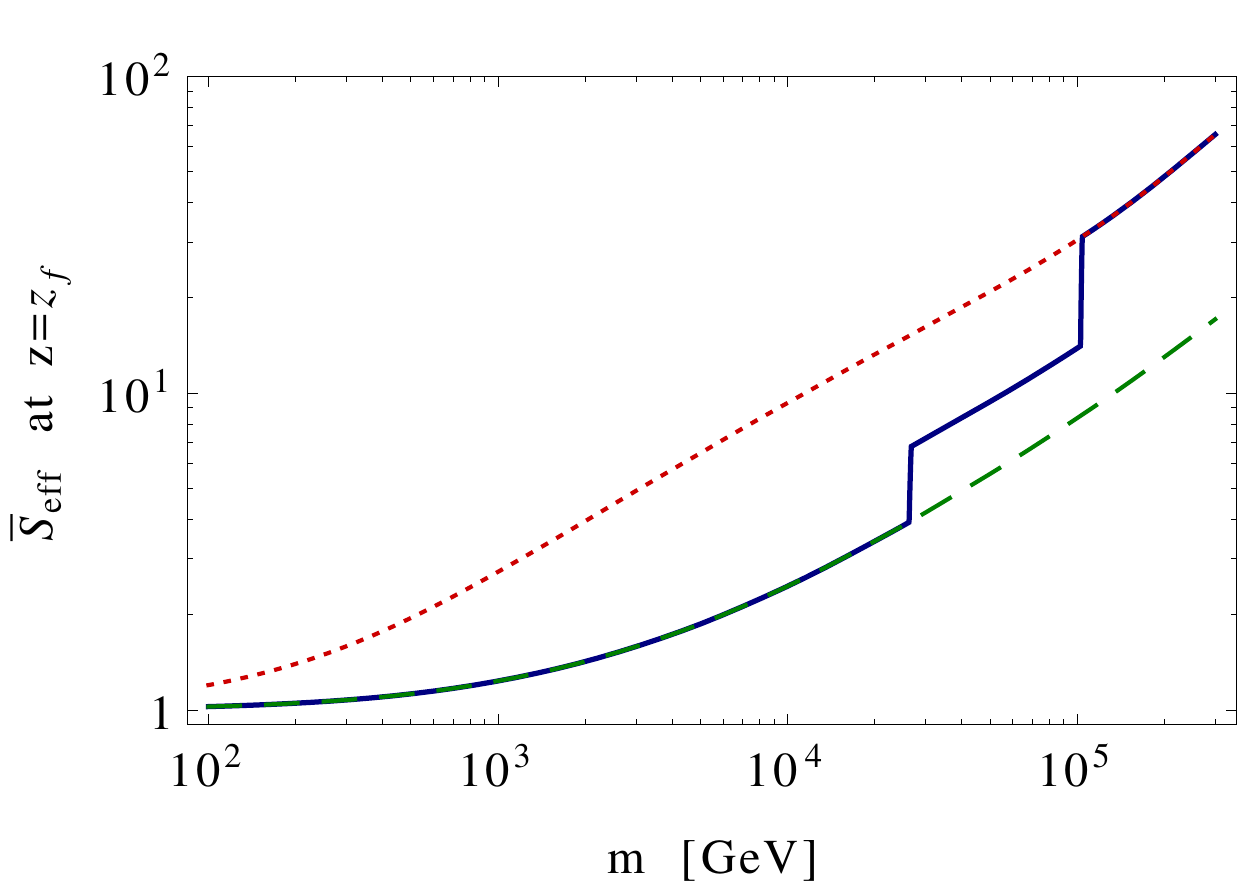}
\caption{
The effective Sommerfeld enhancement, $\bar{S}_{\rm eff}$, evaluated at the time of freeze-out, $z=z_f$, vs. the DM mass $m$ (blue solid line). For $\alpha = \alpha(m)$, we have used the value which yields the observed DM density. Also shown are $\bar{S}_{\rm ann}$ (green dashed) and $\bar{S}_{\rm tot} = \bar{S}_{\rm ann} + \bar{S}_{_{\rm BSF}}$ (red dotted). 
}
\label{fig:Seff FO}
\end{figure}

The relic DM density can also be estimated without employing the coupled differential equations \eqref{eq:Boltz X} -- \eqref{eq:Boltz ortho}, yet incorporating the effect of BSF. BSF contributes effectively to the DM annihilation, provided that the bound states which are formed decay into dark photons faster than the ambient radiation can reionize them into their constituents. Based on the timeline of table~\ref{tab:timeline}, we may thus define an effective thermally averaged SE factor
\begin{equation}    
\bar{S}_{\rm eff} = \left\{
\bal{6}   
&\Sbarann,		& \quad & z \lesssim 0.28&  \\  
&\Sbarann + \frac{\SbarBSF}{4},	& \quad & 0.28 \lesssim  z \ {\rm and} \ c_\a \lesssim f_{\rm ion} (z) \\  
&\Sbarann + \SbarBSF ,	& \quad & f_{\rm ion} (z)  \lesssim c_\a \, .  
\eal  
\right.  
\label{eq:Seff}  
\end{equation}  
We show a plot of $\bar{S}_{\rm eff}$ in Fig.~\ref{fig:Seff FO}. We can then estimate the DM relic abundance from the evolution equation  %
\begin{equation}  
\frac{dY_{_X}}{dz} = - c_1 \frac{\bar{S}_{\rm eff} (z_{_X})}{z^2} Y_{_X}^2   \, ,
\label{eq:Boltz X approx}  
\end{equation}   
for $z \gtrsim z_f$. (We ignore the pair-creation processes, which become unimportant soon after freeze-out.) Equation~\eqref{eq:Boltz X approx} can be analytically integrated to give   
\begin{equation}    
\frac{1}{Y_{_X}(z_s)}  =  \frac{1}{Y_{_X}(z_f)}  + \int_{z_f}^{z_s} dz \: \frac{c_1\, \bar{S}_{\rm eff} (z_{_X})}{z^2}  \,  .  
\label{eq:YX approx}    
\end{equation}    
Using \eq{eq:YX approx}, we find $\a$ as a function of $m$  such that the observed DM density is reproduced. The results are in agreement with those obtained from solving the coupled Boltzmann equations, to better than 2\% accuracy.

\section{Unitarity and critical coupling}

For $\a \simeq 0.54$, the sum $\sann + \sBSF$ becomes equal to the unitarity bound on the inelastic cross-section in \eq{eq:sigma unitarity}. For this $\a$,  the observed DM abundance is reproduced if $m$ equals
\begin{equation} 
m_{\rm uni}^D \simeq 139~\TeV \, . 
\label{eq:m uni Dirac}
\end{equation}
This is the unitarity bound on the mass of thermal relic DM consisting of Dirac fermions. From this, we deduce the corresponding bound on Majorana DM:
\beq 
m_{\rm uni}^M = \sqrt{2}\, m_{\rm uni}^D \simeq 197~\TeV \, .  
\label{eq:m uni Major}
\eeq
The bounds for complex and real scalar DM are the same as for Dirac and Majorana DM respectively. 
Of course, if there is significant entropy release in the universe after DM annihilations become inefficient, or if the dark radiation is at a lower temperature than the SM plasma during freeze-out, the bounds are relaxed accordingly.

The preceding analysis takes into account the belated DM freeze-out and the DM annihilations occurring after that point, due to the SE of $\sann$ and $\sBSF$ at low velocities. The SE results in $m_{\rm uni}$ being larger than what would otherwise be expected.
For comparison, using $\W\DM h^2 \simeq 0.12$~\cite{Ade:2013zuv}, the analysis of Ref.~\cite{Griest:1989wd} would give $m_{\rm uni}^D \simeq  340\, \, \TeV \cdot(\W\DM h^2/2)^{1/2} \simeq 83\, \, \TeV$ for annihilation without SE, where the factor $1/\sqrt{2}$ translates the bound from Majorana to  Dirac DM.  Reference~\cite{Griest:1989wd} also considered the case of annihilation with SE (although they deemed it improbable).  Their estimate for this case would be $m_{\rm uni}^D \simeq 550\, \, \TeV \cdot (\W\DM h^2/2)^{1/2} \simeq 135 \, \, \TeV$, which is close to \eq{eq:m uni Dirac}. (Indeed, close to the unitarity limit, the effect of the ionization of the bound states is negligible.)

\medskip

We may estimate the $\a = \a_{\rm uni}$ for which $\s_{\rm inel} = \s_{\rm inel, \, max}$, under various assumptions for the contributing inelastic processes
\beq
\a_{\rm uni} \simeq \left \{
\bal{6}
&0.86,& \qquad &\s_{\rm inel} = \sann \, ,& \\
&0.54,& \qquad &\s_{\rm inel} = \sann+\sBSF^{(1)} \, ,& \\
&0.47,& \qquad &\s_{\rm inel} = \sann+ \sum_n \sBSF^{(n)} \, .&
\eal
\right.
\eeq
$\a_{\rm uni}$ provides an estimate of the range of validity of the approximation used in evaluating the inelastic cross-section, albeit  not necessarily the most stringent.  According to Gribov~\cite{Gribov:1998kb,Gribov:1999ui}, gauge theories have a critical coupling above which the Coulomb interaction between fermions becomes strong enough to cause a rearrangement of the perturbative vacuum. In QED, this is~\cite{Gribov:1998kb,Gribov:1999ui,Dokshitzer:2004ie,Hoyer:2014gna}  
\beq 
\a_{\rm crit} = \p (1-\sqrt{2/3} )  \simeq 0.58 \, . 
\label{eq:alpha crit} 
\eeq
It is interesting to note that $\a_{\rm crit}$ is close to but larger than $\a_{\rm uni}$ when BSF is taken into account.\footnote{Loop corrections to the direct annihilation and BSF processes (beyond the ladder diagrams considered in Sec.~\ref{sec:rates}), may reduce $\a_{\rm uni}$. However, we expect that these corrections are suppressed by powers of $\a/(4\p)$ which is always small for the range of $\a$ considered here. The same is true if we take into account final states with a larger number of dark photons; these processes are additionally suppressed by phase space.}

\section{Constraints}   

The model and the parametric regime investigated here, as well as other similar scenarios, are viable with respect to observational constraints.  In particular, the coupling of DM to a light or massless particle mediates DM self-interactions.  The most stringent bounds on these interactions arise from the observed ellipticity of Milky-Way-sized haloes. Although significant uncertainties exist, recent simulations and observations estimate the upper bound on the momentum-transfer scattering cross-section per mass of DM to be $\s_{\rm mt}/m \lesssim  2~{\rm barn}/\GeV$~\cite{Peter:2012jh,Rocha:2012jg,Vogelsberger:2012ku,Zavala:2012us}. In the present model,      
\begin{equation}    
\s_{\rm mt} = \int d\W (1-\cos \theta) \frac{d\s_{\rm sc}}{d\W}   = \frac{4\p \a^2}{\mu^2 \vrel^4} \ln [\csc (\th_{\rm min}/2)] \, , \nn 
\end{equation}    
where $\th_{\rm min} > 0$ encodes the effect of screening due to the Debye length in neutral plasma and/or due to a finite mediator mass.  Taking $\ln [\csc (\th_{\rm min}/2)] \sim 10$, and  $\vrel \sim 250~\km/\snd \simeq 8\times 10^{-4}$ for a Milky-Way-sized halo, we estimate     
\begin{equation}  
\frac{\s_{\rm mt}}{m} \approx 0.7~\frac{\rm barn}{\GeV}  \pare{\frac{8 \times 10^{-4}}{\vrel}}^4  \sqpare{\frac{\s_0}{(\s_{\rm ann} \vrel)_c} }    \pare{\frac{\TeV}{m}} \, , \nn  
\end{equation}       
which satisfies the existing constraints in the mass range of interest. (Note that because of the SE of $\sann$ and of the effect of BSF on the DM freeze-out, the observed DM abundance is obtained for $\s_0 < \ang{\sann \vrel}_c$, for $m \gtrsim~\TeV$.) Additional constraints may arise if the dark sector couples to SM particles. Exploring these constraints is beyond the scope of the present work.

\section{Conclusion}

We demonstrated that the formation of bound states in the early universe significantly enhances the annihilation rate of thermal DM with mass above a few TeV, if DM annihilates via a long-range interaction. We argued that this is the only scenario in which thermal DM can be as heavy as unitarity permits. We determined the unitarity bound on the mass of thermal-relic DM to be about 139~TeV for non-self-conjugate DM, and showed that BSF is the dominant annihilation channel in this regime. Importantly, even the weak interactions of the SM manifest as long-range during DM freeze-out, if DM is heavier than a few TeV.

Here we focused on DM consisting of Dirac fermions annihilating into massless dark photons. We determined the dark fine structure constant $\a$ which yields the observed DM abundance, to be up to about 40\% smaller than estimated when BSF is ignored, and up to an order of magnitude smaller than estimated if the Sommerfeld effect is altogether neglected, with the largest discrepancy arising close to the unitarity limit. Our results are presented in Figs.~\ref{fig:alphaVm} and \ref{fig:Omegas}. Our analysis may be extended to other types of DM interacting via a light but massive scalar or vector boson. This is particularly compelling for heavy DM coupled to the weak interactions of the SM, which can be probed at the LHC and future high-energy colliders. It is also important for hidden-sector DM, which may yield observable high-energy astrophysical signals. Indeed, the accurate interpretation of the experimental results necessitates a precise knowledge of the couplings which yield the observed DM abundance.

In fact, the significance of these couplings is even broader. Thermalized species which annihilate more efficiently than these couplings allow, can account for the entirety of DM provided that they carry a particle-antiparticle asymmetry~\cite{Nussinov:1985xr,Davoudiasl:2012uw,Petraki:2013wwa,Zurek:2013wia,Boucenna:2013wba}. Asymmetric DM is motivated by the similarity of the observed dark and ordinary matter abundances~\cite{Nussinov:1985xr}; furthermore, it is a very good host of self-interacting DM~\cite{Petraki:2014uza}, which is favored by observations of the galactic structure~\cite{Spergel:1999mh}.  On the other hand, DM which annihilates less efficiently may have been produced only non-thermally, otherwise it would overclose the universe; candidates in this category are sterile neutrinos~\cite{Dodelson:1993je,Shi:1998km,Petraki:2007gq,Kusenko:2010ik,Canetti:2012vf,Merle:2013gea,Drewes:2013gca}, and axions~\cite{Peccei:1977hh,Kim:1979if,Zhitnitsky:1980tq,Dine:1981rt,Sikivie:2010bq}. The above possibilities arise, of course, within DM theories of very different structure, i.e. very different beyond-SM physics. The precise value of the couplings which produce the observed DM abundance in the symmetric thermal-relic scenario sets the border in the parameter space between structurally different DM theories and is an important quantity for beyond-SM and DM physics even beyond that specific paradigm.

\section*{Acknowledgments}

We thank Alfredo Urbano and Wei Xue for useful conversations. KP thanks Paul Hoyer for illuminating discussions. KP was supported by the Netherlands Foundation for Fundamental Research of Matter (FOM) and the Netherlands Organization for Scientific Research (NWO).

\bibliography{BSF_arXiv_v2.bbl}

\end{document}